\documentclass[]{spie}  %>>> use for US letter paper
\usepackage{amsmath,amsfonts,amssymb}
\usepackage{graphicx}
\usepackage[colorlinks=true, allcolors=blue]{hyperref}

\title{Pioneering high contrast science instruments for planet characterization on giant segmented mirror telescopes}

\author[a,b]{N. Jovanovic}
\author[a,c,d,e]{O. Guyon}
\author[a]{J. Lozi}
\author[e,f,g]{M. Tamura}
\author[h,n]{B. Norris}
\author[h]{P. Tuthill}
\author[i]{E. Huby}
\author[i]{G. Perrin}
\author[i]{S. Lacour}
\author[j]{F. Marchis}
\author[k,l]{G. Duchene}
\author[i]{L. Gauchet}
\author[m]{M. Ireland}
\author[b]{T. Feger}
\author[m]{A. Rains}
\author[m]{J. Bento}
\author[b,n]{C. Schwab}
\author[b]{D. Coutts}
\author[i]{N. Cvetojevic}
\author[b,o]{S. Gross}
\author[b,o]{A. Arriola}
\author[h]{T. Lagadec}
\author[a,p]{S. Goebel}
\author[p]{D. Hall}
\author[p]{S. Jacobson}
\author[q]{B. Mazin}
\author[q]{A. Walter}
\author[q]{J. Massie}
\author[r]{T. Groff}
\author[s]{J. Chilcote}
\author[t]{J. Kasdin}
\author[q]{T. Brandt}
\author[t]{C. Loomis}
\author[t]{M. Galvin}
\author[f]{T. Kotani}
\author[u,v]{H. Kawahara}
\author[w]{F. Martinache}
\author[a]{P. Pathak}
\author[c]{J. Males}

\affil[a]{National Astronomical Observatory of Japan, Subaru Telescope, Hilo, HI, 96720, U.S.A.}
\affil[b]{Department of Physics and Astronomy, Macquarie University, NSW 2109, Australia}
\affil[c]{Steward Observatory, University of Arizona, Tucson, AZ, 85721, U.S.A.}
\affil[d]{College of Optical Sciences, University of Arizona, Tucson, AZ 85721, U.S.A.}
\affil[e]{Astrobiology Center of NINS, 2-21-1, Osawa, Mitaka, Tokyo, 181-8588, Japan}
\affil[f]{National Astronomical Observatory of Japan, 2-21-1 Osawa, Mitaka, Japan}
\affil[g]{The University of Tokyo, Tokyo 113-0033, Japan}
\affil[h]{Sydney Institute for Astronomy (SIfA), Institute for Photonics and Optical Science (IPOS), School of Physics, University of Sydney, NSW 2006, Australia}
\affil[i]{LESIA, Observatoire de Paris, Meudon, 5 Place Jules Janssen, 92195, France}
\affil[j]{Carl Sagan Center at the SETI Institute, Mountain View, CA 94043, USA}
\affil[k]{Astronomy Department, University of California, Berkeley, CA 94720-3411, USA}
\affil[l]{University Grenoble Alpes \& CNRS, Institut de Planetologie et d'Astrophysique de Grenoble (IPAG), Grenoble F-3800, France}
\affil[m]{Research School of Astronomy \& Astrophysics, Australian National University, Canberra ACT 2611, Australia}
\affil[n]{Australian Astronomical Observatory, 105 Delhi Rd, North Ryde NSW 2113, Australia}
\affil[o]{Centre for Ultrahigh bandwidth Devices for Optical Systems (CUDOS)}
\affil[p]{Institute for Astronomy, University of Hawaii, Honolulu, HI 96822, U.S.A.}
\affil[q]{University of California, Department of Physics, Santa Barbara, California 93106, U.S.A.}
\affil[r]{Goddard Space Flight Center, 8800 Greenbelt Rd, Greenbelt, MD 20771}
\affil[s]{Dunlap Institute for Astronomy and Astrophysics, University of Toronto, Toronto, ON M5S 3H4, Canada}
\affil[t]{Princeton University, Department of Mechanical and Aerospace Engineering, Engineering Quadrangle, Olden Street, Princeton,
New Jersey 08544, United States}
\affil[u]{Department of Earth and Planetary Science, The University
of Tokyo, Tokyo 113-0033, Japan}
\affil[v]{Research Center for the Early Universe, School of Science,
The University of Tokyo, Japan}
\affil[w]{Observatoire de la Cote d'Azur, Boulevard de l'Observatoire, Nice, 06304, France}

\authorinfo{Nemanja Jovanovic: E-mail: jovanovic.nem@gmail.com, Telephone: 1 626 395 1214}

\begin{document} 
\maketitle

\begin{abstract}
A suite of science instruments is critical to any high contrast imaging facility, as it defines the science capabilities and observing modes available. SCExAO uses a modular approach which allows for state-of-the-art visitor modules to be tested within an observatory environment on an 8-m class telescope. This allows for rapid prototyping of new and innovative imaging techniques that otherwise take much longer in traditional instrument design. With the aim of maturing science modules for an advanced high contrast imager on an giant segmented mirror telescopes (GSMTs) that will be capable of imaging terrestrial planets, we offer an overview and status update on the various science modules currently under test within the SCExAO instrument. 
\end{abstract}


\section{INTRODUCTION}
\label{sec:intro}  % \label{} allows reference to this section

High contrast imaging is the art of imaging extremely faint structures, such as brown dwarfs, disks and exoplanets in close proximity to bright stars ($<2$" away). This requires the ability to spatially separate the star from the faint structure and then detect the object in the presence of the bright speckle background of the star. The first step to achieving this is to employ advanced adaptive optics (AO) systems known as extreme AO (ExAO) systems. These systems correct for atmospheric perturbations at high temporal and spatial frequencies and can offer $\sim90\%$ Strehl ratios in the near-IR (H-band), in better than median seeing conditions. The past $5$ years has seen numerous ExAO systems come on line to join the already commissioned PALM$3000$~\cite{Dekany2013}, which includes GPI~\cite{Mac2014}, SPHERE~\cite{Vigan2016}, MagAO~\cite{Close2013}, LBTAO~\cite{Esp2010} as well as the Subaru Coronagraphic Extreme Adaptive Optics (SCExAO) instrument~\cite{Nem2015a}.  

The method used to characterize the faint structure/companion is equally as important as the ability to separate and resolve the flux in the first place. For this reason, the scientific capabilities of SCExAO are extremely broad. The VAMPIRES module offers high angular resolution visible light polarimetric imaging with interferometric aperture masking and/or spectral differential imaging at H-alpha and [SII]. This module is optimized for imaging accreting exoplanets, and studying strongly scattering dusty disks and shells around giant stars. The FIRST module exploits spatial filtering combined with pupil remapping interferometry and spectroscopy in a unique way to enhance contrast. It will image close separation binaries and resolved stars and ultimately target hot Jupiters. The RHEA spectrograph offers a 3x3 IFU composed of single mode fibers which is spectrally dispersed with a resolving power of 65000 (600-800 nm). The spatially resolved nature of the IFU allows velocity maps to be created for both disks and the surfaces of giant stars, in which case convection cells could be mapped. In addition, the linewidth of the the Halpha line for an accreting exoplanet could be studied to determine the rate of in flowing material and used to constrain the mass of the growing planet. The GLINT module, uses a photonic chip to null out the star and transmit the signal from the off axis/extended source to a photon counting IR APD (a SAPHIRA detector). With sub-diffraction limited imaging capabilities it will be used to study the dust around giant mass-losing stars, image disks and ultimately RV exoplanets which can not be resolved by other forms of imaging. The SAPHIRA detector is also utilized for sensitive NIR observations as a focal plane array and offers very high frame rates that can freeze the effect of the atmosphere. 

In regards to spectroscopic characterization capabilities of the planet light post-coronagraph, SCExAO offers several modes. Very low resolution will be enabled by the MKIDs detector being developed for SCExAO. This photon counting imager will be used for focal plane wavefront control but also allows for the possibility for high temporal frequency data collection. Low resolution is provided by the recently commissioned CHARIS integral field spectrograph, which has replaced the HiCIAO imager. HiCIAO originally enabled differential narrow-band imaging with polarimetry over a wider field ($\sim20$"). CHARIS on the other hand offers a unique broad band mode which can image from J-K band in a single shot providing unprecedented discrimination between speckles for planet discovery. High-resolution spectroscopy will be provided by injecting light into the IR Doppler instrument (y-H band, R~65000), purpose built for an RV survey of nearby Mdwarfs. This capability will expand the SCExAO tool kit and allow complementary spectra at both low and high resolution to be collected for the first time.

These advanced imaging capabilities are continuously evolving and the aim is to pioneer and prototype them in preparation for an optimized GSMT instrument. In this paper we present the modules of SCExAO, their capabilities, status, and the exciting science to be generated thus far. We describe how these capabilities will play an integral role for an advanced GSMT imager.

\section{Instrument overview}
\begin{figure*}
\centering 
\includegraphics[width=0.95\linewidth]{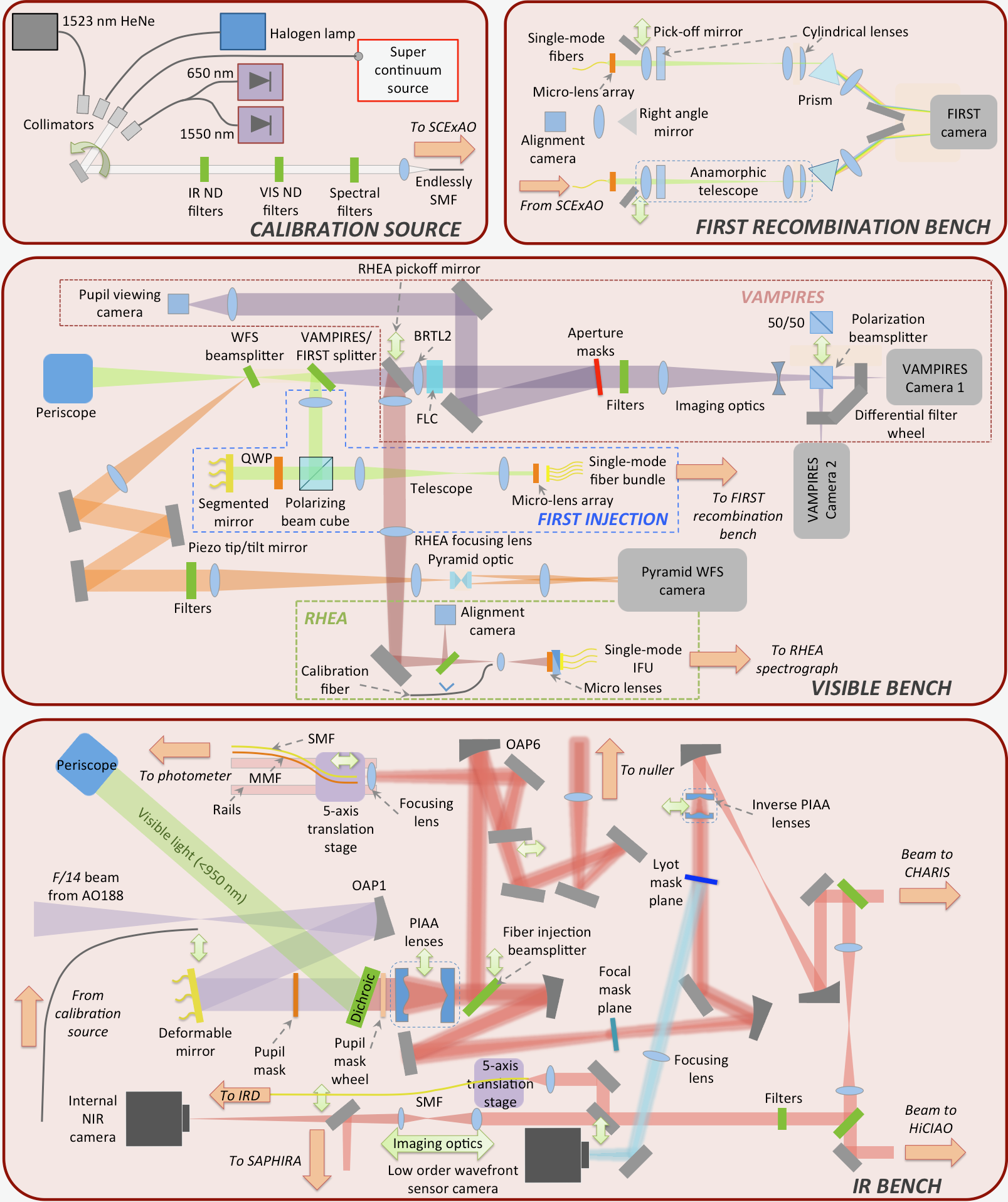}
\caption{\footnotesize Schematic of the SCExAO instrument. (Top left) The calibration source which allows the user to select between a super continuum source, 1523 nm HeNe laser, $650/1550$~nm laser diode and a halogen lamp. (Top right) The photometric setup for calibrating the coupling efficiency. (Middle) Visible bench of SCExAO which hosts the pyramid wavefront sensor. (Bottom) Infrared bench of SCExAO which hosts the fiber injection, the deformable mirror and the PIAA lenses. Dual head green arrows indicate that a given optic can be translated in/out of or along the beam. Orange arrows indicate light entering or leaving the designated bench at that location.}
\label{fig:schematic}
\end{figure*}

A schematic of the SCExAO instrument as of May $2017$ is shown in Fig.~\ref{fig:schematic}. A comprehensive overview of the instrument can be found in Jovanovic \textit{et. al.} ($2015$), so here we offer a basic overview and focus mainly on the science instruments. In addition, we refer the reader for specific details and results about each module to the numerous other articles in SPIE proceedings~\cite{Lozi2016a,Lozi2016b,Pathak2016a,Martinache2016a,Goebel2016,Norris2016a,Huby2016,Rains2016}. 

SCExAO operates downstream of the facility AO system, AO$188$~\cite{Min2010} at the Subaru Telescope. AO$188$ offers Strehl ratios of $30-40\%$ in the H-band, in median seeing conditions. The corrected beam is injected into the bottom bench of SCExAO. After bouncing off of the $2000$-element deformable mirror (DM, Boston MicroMachines Corporation) the beam is split. Wavelengths shorter than $<950$~nm are directed towards a periscope and routed to the visible bench. Longer wavelength light continues to propagate on the bottom bench (IR bench). 

Light between $800$-$950$~nm is fed to the pyramid wavefront sensor (PyWFS). This is the high order wavefront sensor SCExAO uses to reach ExAO levels of performance. This first stage of wavefront correction provided by AO$188$ is needed to make sure that the beam delivered to the PyWFS is within the linear range of the sensor. The PyWFS is used to drive the ExAO loop because of its exquisite sensitivity to high order modes and operates at $1.5$--$3.5$~kHz with very low latency ($\sim1$~ms)~\cite{Lozi2016a,Lozi2016b}. The left over light on the visible bench is used for scientific characterization by the VAMPIRES, FIRST and RHEA modules discussed below. 

The light transmitted by the dichroic on the IR bench can be used for coronagraphy or interferometric nulling. In this case a number of coronagraphs including the Phase Induced Amplitude Apodization (PIAA)~\cite{Guyon2003} and the vector vortex~\cite{Mawet2010,Kuhn2016} versions are available, to name a few. These coronagraphs are optimized for small inner working angle ($1$-$3~\lambda/D$) operation in the near-IR (J-K bands). Post-coronagraph the light is fed to the recently commissioned facility imager, CHARIS~\cite{Brandt2014,Groff2016}. 

The following section summarizes the core science modules of SCExAO, their status and recent results. 

\section{The science modules: overview and status}
\subsection{Visible Aperture Masking Polarimetric Interferometer for Resolving Exoplanetory Signatures}
VAMPIRES is an aperture masking inteferometer providing sub-diffraction limited spatial resolution images with polarimetry, and is commissioned and operational~\cite{Norris2015,Norris2016a}. It operates between $600$--$800$~nm with $5$ discrete narrowband filters ($50$~nm bandwidth) spread across this range. It is optimized to detect structures that are strongly scattering, such as dusty shells around giant stars or disks around young stars. These create a polarized signature that VAMPIRES is ideally suited to detect. Operating at short wavelengths, the interferometer offers very high spatial resolution (approaching $10$~mas) imaging which when combined with the range of spectral filters enables dusty structures to be studied with high fidelity and the nature of the scatterers to be constrained.    

VAMPIRES has recently undergone an upgrade, which includes the installation of a faster polarization switching element and the use of two EMCCD cameras (one for each polarization). Since only a fraction of each detector is used for typical observations, they can be operated in cropped mode allowing for maximum imaging speeds (up to $500$~Hz) to freeze the turbulence. In addition to the improvement in speed, VAMPIRES has been upgraded to enable spectral differential imaging at H-alpha and several other wavelengths of interest ([SII] for example). This was motivated by the recent discovery of an accreting exoplanet directly imaged at $656$~nm~\cite{Sallum2015}. This mode is used with a 50/50 intensity beam splitter in place of the polarization beam splitter and independent of the polarimetric imaging mode currently. It can be operated with aperture masks as well as in full pupil imaging mode.   

The rebuild is mostly complete with some final focal plane masks still outstanding. The dual camera polarimetric mode has been validated and is back to smooth operation. Preliminary tests have been conducted with the spectral differential imaging mode. 

Now that SCExAO has reached the ExAO regime and frequently operates above $80\%$ Strehl ratio in the H-band, VAMPIRES often achieves a diffraction-limited core at the shorter wavelengths as well. The polarimetric mode has been used to reveal an asymmetric dust shell around mu Cephius and study the disk at unprecedented spatial scales of AB Aur. The data of these results are still under analysis.   

\subsection{Fibered Imager foR a Single Telescope}
FIRST is a pupil-remapping interferometer. Similar to VAMPIRES, it interferes sub-sections of the pupil-plane together, to calibrate the wavefront and conduct sub-diffraction limited imaging. However, unlike VAMPIRES it collects the light from the sub-pupils into single-mode fibers with the aid of a micro-lens array that can then be reformatted to a non-redundant output array, optimized for down stream interferometric image formation. The remapping enables the light from the entire input pupil to be utilized at once: increasing signal and Fourier coverage. Since the fibers are reformatted to a linear slit, they can be cross-dispersed as well offering continuous spectral information from $600$--$800$~nm with a resolving power of $R=100$~\cite{Huby2016}.  

This imager is still undergoing commissioning. Although most of the hardware has been installed, new control software needs to be implemented for efficient alignment and data collection both in the laboratory and on-sky. Once these upgrades are made, FIRST will be used to study very close separation binaries as well as the surfaces of giant stars. In one to two years, the recombination bench will receive an upgrade which will see the classical Fizeau beam combiner replaced with an integrated photonic version instead. The elements are currently being developed for this upgrade.     

\subsection{The Replicable High-resolution Exoplanet and Astroseismology spectrograph}\label{s:RHEA}
\begin{figure*}[hb!]
\centering 
\includegraphics[width=0.80\linewidth]{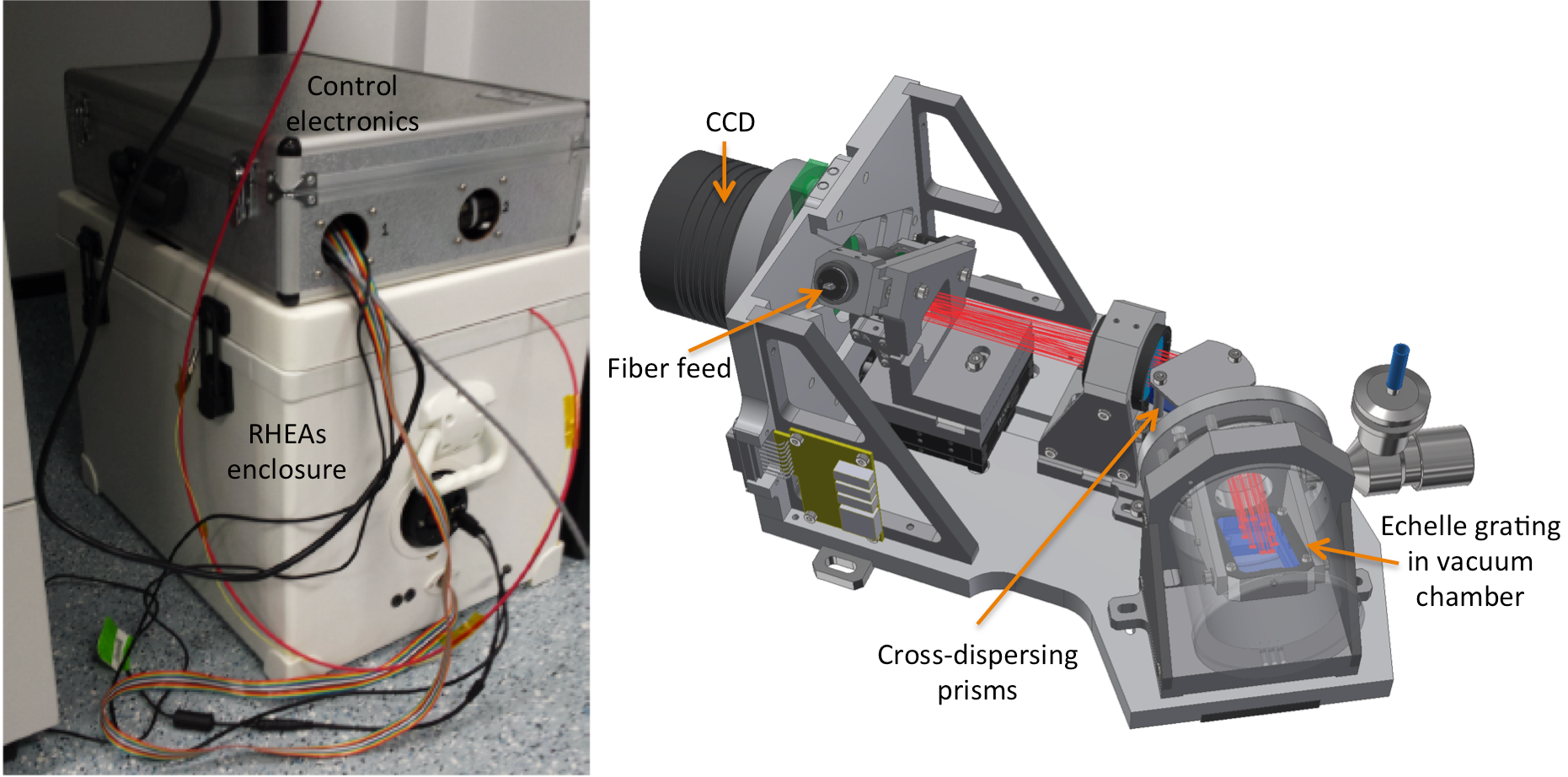}
\caption{\footnotesize (Left) The RHEA spectrograph inside an icebox for thermal stabilization. Control electronics box can be seen on top. (Right) The RHEA spectrograph deployed at Subaru Telescope.}
\label{fig:rhea}
\end{figure*}

The RHEA spectrograph is fed by a single-mode fiber integral field unit (IFU). The IFU consists of a $3\times3$ array of $1$~mm lenslets feeding single-mode fibers and operates from $600$--$800$~nm. The projected on-sky separation of the fibers is $16$~mas. The $9$ fibers are injected into an Echelle spectrograph, which offers a resolving power of $\sim60000$. RHEA is very small (see Fig.~\ref{fig:rhea}) owing to the fact it operates in the diffraction limit, occupying a footprint of less than $30\times30\times30$~cm$^{3}$. The R$2$ Echelle grating and spectrograph as a whole are easily stabilized with very simple temperature control. This means the instrument can achieve an intrinsic precision of $\sim1$~m/s over a night~\cite{feger2016a, feger2016b}. 

The high spatial and spectral resolution make it ideal for mapping the surfaces of giant stars in order to study convection and stellar rotation for the first time, study resolved protoplanetary disk structures as well as for the detailed characterization of accreting protoplanets~\cite{Sallum2015}. An instrument like RHEA behind SCExAO could potentially shed light on the accretion rate and in-flow velocity, which would allow the mass and density of the forming exoplanet to be constrained. This data combined with the spatial location in the protoplanetary disk will aid in refining planetary formation models which are currently unconstrained.

The fiber injection for RHEA is located on the visible bench. To center the IFU on a star, an alignment camera is used. This camera looks at the star and a beam which is counter-propagated from the spectrograph to the IFU and steered onto the camera via a retro-reflecting cube simultaneously. By translating the IFU laterally, it is possible to overlap the reverse injected beam with the star which will ensure that the star is aligned with the IFU and will be coupled in. The injection and spectrograph have been installed on SCExAO and are currently undergoing commissioning~\cite{Rains2016}. After the original installation we observed fringing in the spectra owing to the fact that the $9$-beams were partially overlapped within a single order, on the detector, which was a result of the $10$~m fibers being too similar in length (i.e. within the relatively long coherence length for each narrow spectral channel). This was remedied with the construction of a fiber bundle with dissimilar fiber lengths, which also enhanced throughput. 

RHEA has been tested on-sky on both point sources and resolved stars. The total throughput from sky to detector was determined to be a factor of $5$--$10$ below expectation. Experiments are on-going to determine which element in the optical chain is responsible for the throughput loss. The spectrograph optics have been thoroughly tested and it was concluded that they were not the source of this loss. 

In parallel, continuous improvements in the performance of the AO system have addressed the other major roadblock to this modules success, namely temporally varying low order aberrations. The aim is to have the RHEA module open for science use in the next 6 months.

\subsection{Nulling interferometry}
A nulling interferometer takes beams from multiple telescopes, or in this case, multiple points of a single telescope pupil and combines them interferometrically just like an aperture mask, for example. In the case of a nuller however, the aim is to look for faint objects in the dark fringe, or so-called 'null': while the object moves through the field (assumes fixed pupil observations). This is an interferometric analog to a coronagraph. This can be done in a photonic coupler/splitter. If the phase of one of the arms is carefully controlled at the input to the coupler it is possible to get destructive interference at one port and all the light sent to second port. The light that has been suppressed is from the host star and since the faint companion is incoherent with this, it still continues to leak out the dark port of the coupler where is can be detected. In this way, the photon-noise from the star can be reduced and the detectability of the faint companion improved. GLINT is a nuller based on a $3$D laser written photonic chip that is currently being commissioned at Subaru~\cite{Norris2016b}. The system only has $2$ ports (i.e. interferes two sub-pupils) thus far but upgrades to take light from numerous parts of the pupil are planned for the future. In addition, the eventual goal is to disperse the near-IR (J and H-band) outputs of the nulling chip onto a read-out noise free detector that can run at high speed to keep up with the atmosphere, so that the null can be tracked at all wavelengths at once.  

A key feature of the photonic nulling chip is that it allows photometric channels and anti-null outputs, in addition to the actual null output, to be measured simultaneously (in other methods non-simultaneous measurements via chopping must be used). This enables a very accurate implementation of the numerical-self-calibration technique (NSC)~\cite{hanot2011}, in which the diversity of phase errors induced by residual seeing after AO correction is actually used to probe the null depth. A statistical analysis on the multiple measurements is conducted to produce accurate null measurements without the need for a separate calibrator star, essentially turning the disadvantage of residual seeing into an advantage.

The module is still undergoing commissioning but has already demonstrated basic operation on an extended object (larger dusty star) as well as a binary star system. The results however were hampered by the low-wind effect as well as telescope vibrations. Efforts on the wavefront control side are ongoing to try to remedy these issues. Once resolved, the system will be characterized in the ExAO regime of operation. The eventual goal will be to use the photonic nuller to image RV planets too close to the host star for traditional coronagraphic techniques. The sub-diffraction limited capabilities of GLINT will open up a new regime of exoplanet imaging once fully operational.      

\subsection{SAPHIRA}
The SAPHIRA array manufactured by Leonardo, comprises $320\times256$ avalanche photodiodes (APDs) made of HgCdTe. This array can offer sub-electron read noise when gain is used in conjunction with up-the-ramp/Fowler readout techniques. This sensitivity will be critical for imaging faint structures. Courtesy of the low noise, these arrays can be operated at kHz speeds enabling for real time focal or pupil plane wavefront control loops to be driven in the near-IR for the first time. In addition, the high frame rates enable the opportunity for speckle or lucky imaging in the near-IR.    

The array has been deployed to SCExAO for the last few years. The imager has faced two large hurdles thus far. The first is noise. The dominant noise source has been radio frequency interference (RFI) noise. By eliminating ground loops, and employing careful shielding this has been mitigated to a large extent but is still far from ideal. The second major hurdle has been the development of the so called ``Pizza Box" read out electronics. These are being developed by the IfA in Hilo, Hawaii. They are based on FPGAs and offer MHz rate pixel read outs and can be connected to a computer via low latency USB 3.0. The electronics were recently commissioned and now the camera deployed on SCExAO can offer $1.6$~kHz frame rate imaging with a $128\times128$~pixel field-of-view ($\sim1.6\times1.6$" on-sky) in the H-band. Other sub-window modes are also now in routine operation. 

It is currently being used to study the lifetime of speckles in the near-IR so future wavefront control systems in these bands can be optimized~\cite{Goebel2016}. In addition, it has been used to conduct speckle imaging of giant resolved stars.   

\subsection{The Microwave Kinetic Inductance Detector Exoplanet Camera}
SCExAO will also receive a dedicated Microwave Kinetic Inductance Detector (MKID) array named the MKID Exoplanet Camera (MEC). The array will consist of $20$~kpixels, the largest built to date and be optimized for efficient operation between $800$-$1400$~nm. The MKID array will offer read-noise-free measurements at kHz speeds. In addition, MKID arrays offer energy discrimination. Recently, PtSi arrays were successfully realized, which allows for a resolving power between $8$--$10$ in the J-band.     

The primary motivation for such an array is to conduct focal plane wavefront sensing, specifically designed to target long lived, chromatic speckle residuals~\cite{Guyon2016}. It is projected that it will improve the contrast by $1$-$2$ orders of magnitude close in to the host star. After the MKID array has helped dig a dark region it will be used as a very low resolution spectrograph to characterize the faint structure/companion around the host. Besides offering spectral discrimination between the speckles and the structure/companion, it will also discriminate between the two based on their statistical properties. This new form of discrimination has never been demonstrated on-sky before and will be a powerful tool.    

MEC is undergoing final assembly and testing before shipping to Subaru Telescope in mid-December $2017$.

\subsection{Coronagraphic High Angular Resolution Imaging Spectrograph}
\begin{figure*}[ht!]
\centering 
\includegraphics[width=0.95\linewidth]{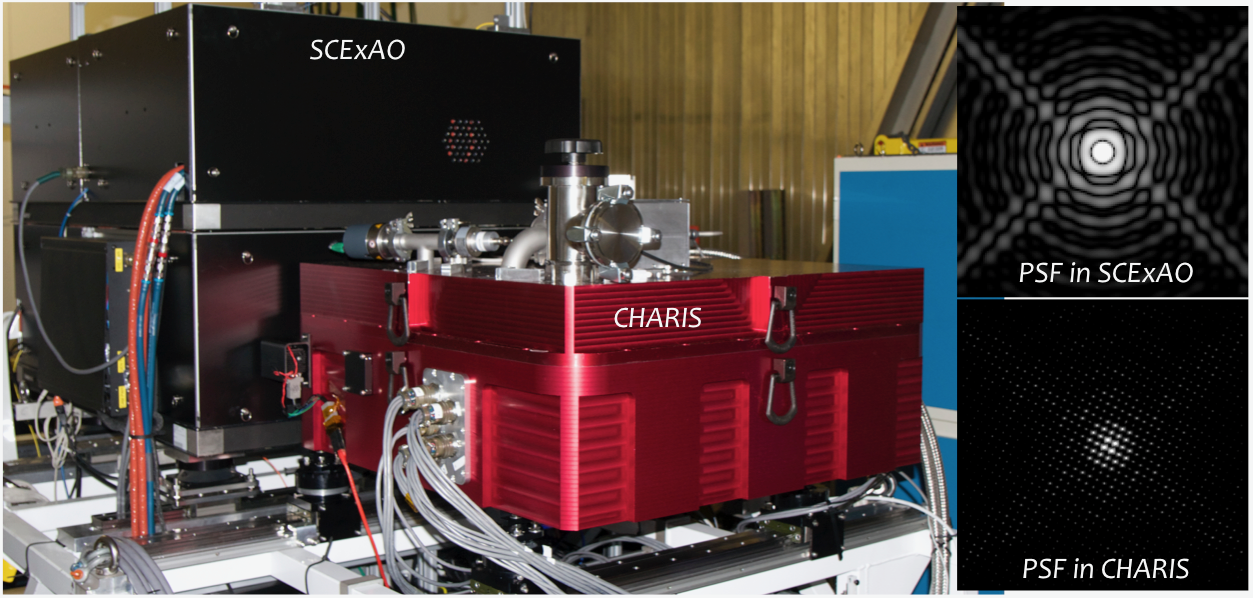}
\caption{\footnotesize (Left) An image of the CHARIS instrument aligned behind the SCExAO instrument. (Right) The point-spread function (PSF) inside SCExAO and after the micro-lens array in CHARIS.}
\label{fig:charis}
\end{figure*}

The integral field spectrograph, CHARIS (shown in fig.~\ref{fig:charis}), has now replaced the HiCIAO imager. A micro-lens array is used to section the focal plane into 'PSFlets' which are dispersed via a prism onto a H$2$RG detector (see fig.~\ref{fig:charis}). CHARIS offers a unique low resolution mode, $R\sim20$, from J-K band in a single shot and a high resolution mode, $R\sim70$--$90$, mode in any single band at a time~\cite{Groff2016}. CHARIS provides spatially resolved spectral information that will be used to study the atmospheres of substellar companions, and differentiate the companion from speckles and background objects. 

It has recently been commissioned  and is fully operational. The low resolution mode is widely used owing to the powerful spectral discrimination offered between the speckles and the companions/disk. A data reduction pipeline has been implemented which offers in addition to aperture photometry and optimal extraction, chi-squared fitting of template spectra for enhanced spectral extraction~\cite{brandt2017}. The system has already delivered high quality spectra for a range of targets including HR8799, Kappa And, LkCa15, HD1160 to name a few. 

A polarimetric mode will be implemented over the next year, which will compliment the previous HiCIAO PDI mode~\cite{Yang2017} that was successful from $0.5$--$5$" but at much smaller angular separations ($<0.5$"). In addition SCExAO is currently commissioning several new broadband coronagraphs including the vector apodized phase plate (vAPP). This coronagraph was optimized to deliver a dark hole at a contrast level of $10^{-5}$ from $2$--$12~\lambda/D$ across J-K bands. When commissioned this will push SCExAO/CHARIS to very small inner working angles, a regime that has not been explored thus far.      

\subsection{High dispersion coronography}
In addition to the low spectral resolution offered by MEC and CHARIS, SCExAO can feed light, post-coronagraph to the IR Doppler (IRD) instrument at Subaru Telescope~\cite{Kotani2014}. IRD is a high resolution, $R\sim65000$ spectrograph that operates from y-H band. It is well stabilized in regards to both temperature and flexure and will be used for an RV survey around M-stars. The aim is to achieve a precision of $1$~m/s on brighter targets, enough to detect exo-Earths. This precision radial velocity machine is also ideal for characterizing exoplanet atmospheres.  

An injection unit has been developed post-coronagraph (near the LOWFS in Fig.~\ref{fig:schematic}), that allows for a fiber to be positioned on the location of a known planet in the field and route it to IRD. A single-mode fiber will be used to enhance the suppression of stellar photons coupled into the fiber. Speckle control will be performed on the location of the fiber to reduce the photon noise further~\cite{mawet2017}. By reducing the photon-noise it is possible to integrate longer on the planet light to overcome detector noise. This mode will offer high resolution spectroscopic insights into the chemical composition of the brightest substellar companions and has never been attempted thus far. 

IRD is currently undergoing commissioning and will begin routine science operation in early $2018$. This mode will then need to be commissioned. Key items that need development before then include tracking the moving planet in the field with the fiber, speckle control at a single spatial frequency based on monitoring the light through the fiber and exploiting predictive control to chase the turbulent speckles in the field. This mode will offer the first high resolution ($R>10000$) spectra of exoplanet atmospheres to date.

\section{Summary}
The SCExAO instrument employees numerous innovative science machines for the characterization of exoplanet and brown dwarf atmospheres and the detailed study of disks. These instruments cover a vast range of the spectrum (from $600$--$2400$~nm), bandwidths (from $<1$--$100$~nm) and temporal ranges (micro-seconds to minutes). These modules, along with new concepts will continue to be developed within the SCExAO project in order to fast track an optimized instrument for a GSMT capable of studying terrestrial planets. 

Owing to the testbed nature of SCExAO, it would be possible to upgrade the instrument to form the basis for a first light imager for a GSMT. The upgrades required would include the addition of a high actuator count, high stroke DM upstream of SCExAO to replace AO188 so that ExAO performance could be achieved on the larger apertures. Other upgrades include adjusting the input optics to accept the beam from a GSMT, considerations for the mounting at the new telescope and more seriously, tying in the SCExAO wavefront control system to utilize the adaptive secondary at the observatory. These modifications will enable the realization of a high contrast imager that could be used for preliminary science close to first light. More importantly, the testbed would offer critical insights into how the telescope operates and any issues that would need to be considered when developing an optimized instrument down the track.

\acknowledgments % equivalent to \section*{ACKNOWLEDGMENTS}       
 
The authors acknowledge support from the JSPS (Grant-in-Aid for Research \#$23340051$, \#$26220704$ \& \#$15$H$02063$). This work was supported by the Astrobiology Center (ABC) of the National Institutes of Natural Sciences, Japan and the directors contingency fund at Subaru Telescope. The authors wish to recognize and acknowledge the very significant cultural role and reverence that the summit of Maunakea has always had within the indigenous Hawaiian community. We are most fortunate to have the opportunity to conduct observations from this mountain. 

% References
\bibliography{report} % bibliography data in report.bib
\bibliographystyle{spiebib} % makes bibtex use spiebib.bst

\end{document}